\documentclass{vgtc}                          

\graphicspath{{figures/}{pictures/}{images/}{./}} 

\usepackage{times}                     

\usepackage{tabu}                      
\usepackage{booktabs}                  
\usepackage{lipsum}                    
\usepackage{mwe}                       
\usepackage{mathptmx}                  
\usepackage{color,soul}
\usepackage{graphicx}
\usepackage{subcaption} 

\onlineid{0}

\vgtccategory{Research}

\vgtcinsertpkg

\title{
Steering LLM Summarization with Visual Workspaces for Sensemaking
}

\author{Xuxin Tang\thanks{e-mail: xuxintang@vt.edu}\\ %
        \scriptsize Virginia Tech %
\and Eric Krokos\\%
    \scriptsize US Department of Defense
\and Can Liu\thanks{e-mail: canliu@cityu.edu.hk}\\ %
     \scriptsize City University of Hong Kong %
\and Kylie Davidson\thanks{e-mail: kyliedavidson@vt.edu}\\%
    \scriptsize Virginia Tech %
\and Kirsten Whitley\\ %
    \scriptsize US Department of Defense
\and Naren Ramakrishnan \thanks{e-mail: naren@cs.vt.edu}\\%
    \scriptsize Virginia Tech %
\and Chris\ North \thanks{e-mail: north@cs.vt.edu}\\%
    \scriptsize Virginia Tech %
     }

\abstract{

   Large Language Models (LLMs) have been widely applied in summarization due to 
   their speedy and high-quality text generation. Summarization for sensemaking involves information compression and insight extraction. 
   Human guidance in sensemaking tasks can prioritize and cluster relevant information for LLMs. 
   However, users must translate their cognitive thinking into natural language to communicate with LLMs. Can we use more readable and operable visual representations to guide the summarization process for sensemaking?
   Therefore, we propose introducing an intermediate step--a schematic visual workspace for human sensemaking--before the LLM generation to steer and refine the summarization process. We conduct a series of proof-of-concept experiments to investigate the potential for enhancing the summarization by GPT-4 through visual workspaces. Leveraging a textual sensemaking dataset with a ground truth summary, we evaluate the impact of a human-generated visual workspace on LLM-generated summarization of the dataset and assess the effectiveness of space-steered summarization. We categorize several types of extractable information from typical human workspaces that can be injected into engineered prompts to steer the LLM summarization. The results demonstrate how such workspaces can help align an LLM with the ground truth, leading to more accurate summarization results than without the workspaces.

} 


\begin{document}


\firstsection{Introduction}

\maketitle
Summarization involves creating a condensed version of one or more documents by expressing the most important facts or ideas. Given multiple documents, summarization needs to identify connections within vast, unstructured, and dispersed information and perform a modicum of distillation and sensemaking to extract meaningful insights for reporting \cite{model2006making, pirolli2005sensemaking}. While human analysts are good at schematizing key insights, writing such a textual summary is complex and time-consuming.

LLMs have been widely applied in summarization due to their generation capabilities. However, current interfaces for interacting with LLMs generally support conversational interaction, necessitating that humans translate their cognitive thinking into natural language.
When handling multi-document sensemaking, human analysts often use spatial interfaces that support visual schematics to facilitate analysis and summarization. Can such readable and operable visual workspaces be used to augment LLM summarization to achieve better results?

Previous research on LLM summarization has primarily focused on issues like hallucination \cite{tang2024tofueval}, similarity with the original dataset, and quality \cite{gao2023human} in open-ended tasks. We lack an understanding of how accurately LLMs can summarize when analyzing multiple given documents in sensemaking tasks. Can we leverage publicly available datasets with established ground truths from visual analytics to evaluate the outcomes of LLM summarization for sensemaking?

To address these challenges, we hypothesize that using an intermediate workspace, such as Space to Think \cite{andrews2010space}, as a preliminary step before LLM summarization can help improve the summarization process. This workspace should facilitate human analytics and enhance LLMs' reasoning and summarization capabilities. To investigate this hypothesis, we leverage a complex sensemaking dataset with ground truth to evaluate the summarization performance, both with or without the intermediate workspace. 

Our contribution includes: 1) proposing the concept of space-steered summarization using interactive visual workspaces to guide LLM summarization, 2) transforming visual representations in the intermediate workspace into prompts for LLM summarization; 3) conducting replicated experiments to evaluate the efficiency of the space-steered summarization.
Our findings demonstrate that steering LLM summarization through visual workspaces can generate results highly aligned with the ground truth compared to those without the workspaces. It indicates the significant potential of the intermediate workspaces as a critical stage for improving human-AI collaboration, particularly in sensemaking, where humans collaborate with AI systems to interpret, understand, and make sense of complex information. 


\section{Background of Summarization for Sensemaking}
\subsection{Human Summarization}
Making sense of multiple documents is an intricate and demanding cognitive process \cite{ancona2012framing,pirolli2005sensemaking,stasko2007jigsaw,weick2005organizing}. This process involves organizing data to structure the unknown, helping individuals gain a deeper and more nuanced comprehension by extrapolating information from the data \cite{ancona2012framing}. Pirolli \& Card have delineated the different phases of sensemaking and their cyclical integration, which enhances an analyst's comprehension of data \cite{pirolli2005sensemaking}. The two principal cycles, foraging and sensemaking, require organization and schematization of the data to facilitate the sensemaking process.
Such schematization typically involves introducing an intermediate step using interactive spatial representations of the relevant data. In response, visual workspaces such as Space to Think \cite{andrews2010space} and Jigsaw \cite{stasko2007jigsaw} were proposed as intermediate representations to facilitate human sensemaking and analytics.

\begin{figure}[htb]
 \centering 
 \includegraphics[width=0.8\columnwidth]{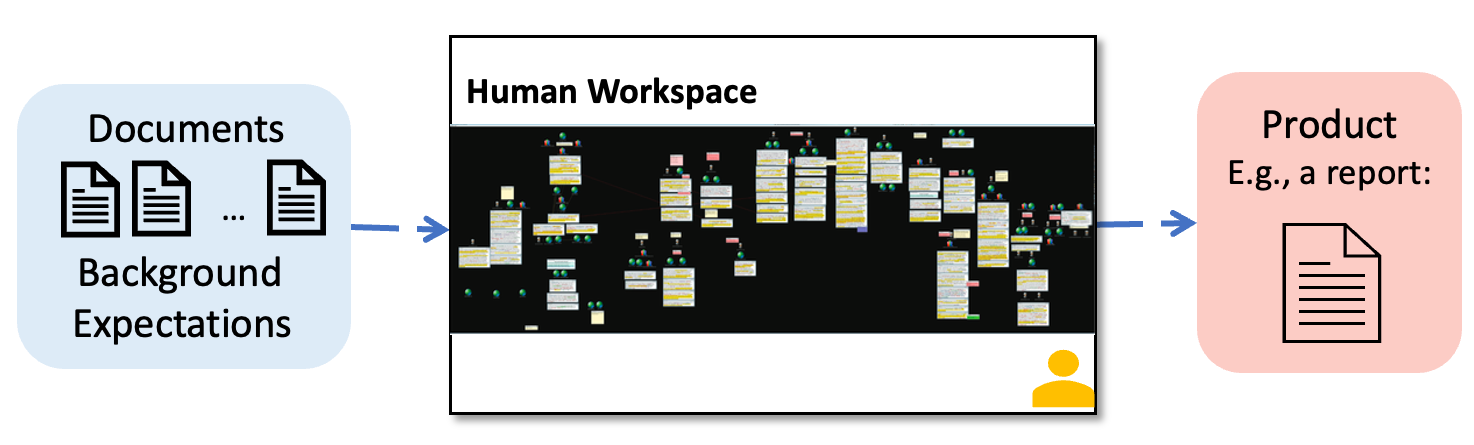}
 \caption{Human summarization for sensemaking. We take Space to Think \cite{andrews2010space} as an example for the intermediate workspace.}
 \label{fig:hm_sense}
\end{figure}

As the final step of the sensemaking process, humans need to extract and summarize the most important facts, insights, and takeaways from those intermediate visual representations into a final product for presentation, typically in the form of a written report, as shown in Figure \ref{fig:hm_sense}.

\subsection{LLM Summarization}
\begin{figure}[htb]
 \centering 
 \includegraphics[width=0.8\columnwidth]{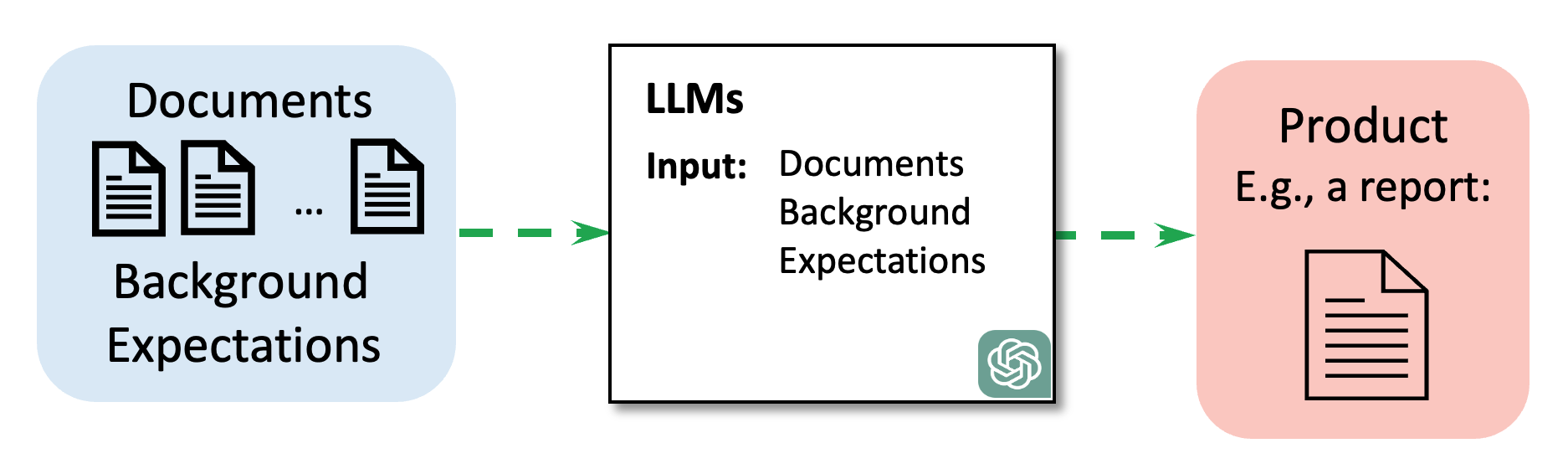}
 \caption{LLM summarization for sensemaking.}
 \label{fig:llm_sense}
\end{figure}

Recent advancements in Large Language Models (LLMs), such as GPT-4 \cite{openai2023gpt}, have showcased their remarkable capability to synthesize and summarize insights from vast datasets \cite{bubeck2023sparks}. The summarization by LLMs is shown in Figure \ref{fig:llm_sense}: LLMs are provided with all the information of the given sensemaking task, then generate the final answer as the product.

However, analysts must translate their mental models into natural language during each interaction to guide the LLM summarization for sensemaking, such as saying, ``\textit{moving document 7 out of cluster 1 and generate the report again}". Is it feasible to make the intermediate space readable by LLMs so that simple interactions such as dragging and clicking can achieve the same effect? Therefore, our method introduces visual workspaces, which are more operable for humans, to steer the LLM summarization.


\section{Space-steered Summarization}
\subsection{Space to Think}
Space-steered summarization is inspired by the ``\textit{Space to Think}" concept introduced by Andrews et al. \cite{andrews2010space}.
Based on psychological theories of distributed and embodied cognition \cite{wilson2002six}, they identified two primary functions of an intermediate workspace for human analysis and sensemaking:
\begin{enumerate}
    \item \textbf{External memory} enables analysts to externalize and offload their cognitive process into the space for later rapid visual retrieval. The memory function of ``\textit{Space to Think}" can provide context for users at multiple levels: 1) highlighted texts offer rapid access to context within a document, and 2) a document in space can be contextualized by its neighbors.
    \item \textbf{Semantic layer} enables analysts to add structural information that supports synthesis. Another critical role played by space is to offer a flexible semantic layer that adds meaning to the displayed information, representing many spatial relationships \cite{kirsh1995intelligent} and annotations and serving as a medium for creating complex structures like clusters \cite{shipman1995finding}.
\end{enumerate}

\subsection{Space-steered Summarization}
\begin{figure}[htb]
 \centering 
 \includegraphics[width=\columnwidth]{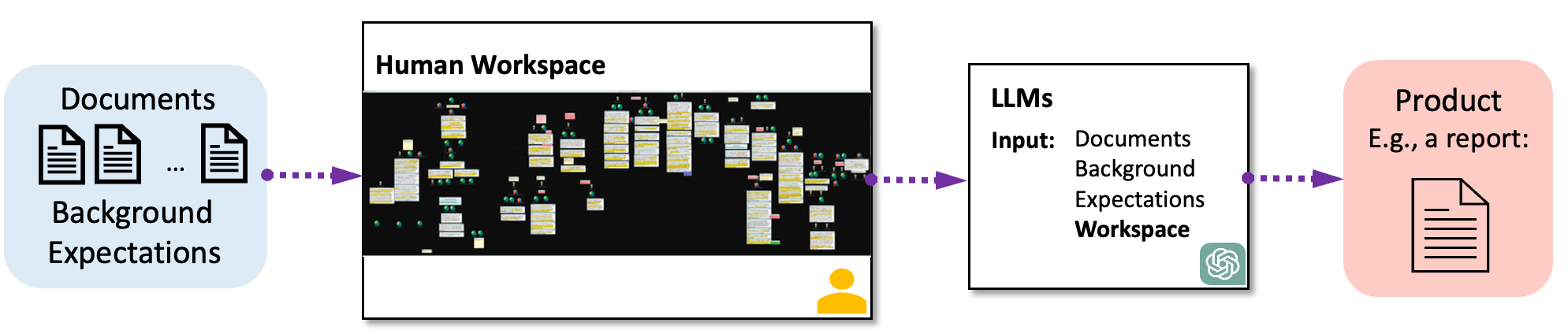}
 \caption{Space-steered summarization process.}
 \label{fig:human_steer_ai}
\end{figure}

The space-steered summarization process, illustrated in Figure \ref{fig:human_steer_ai}, positions LLM summarization as the subsequent step following human workspace construction. To steer LLM summarization, the prompt includes the workspace, background description, and output expectation.

The two primary functions of ``\textit{Space to Think}" are highly significant for LLM summarization by serving as additional context:
\begin{enumerate}

\item \textbf{External memory} enables LLMs to remember facts, preferences, or previous interactions \cite{zhang2024survey}. Recent lessons \cite{lewis2020retrieval, memaugLLM2023, zhong2022training} underscore the significance of external memory for LLMs. 
This memory can be updated and accessed dynamically, enhancing the model's ability to provide personalized responses or maintain continuity in long conversations. Using intermediate workspace as external memory for LLMs allows for efficient retrieval of human-extracted information.

\item In previous research, structural information in the \textbf{semantic layer} empowered LLMs in text generation by customizing text structure \cite{zhang2023visar} and enhancing knowledge retrieval \cite{feng2023knowledge}. However, structural information serves a broader purpose. It reflects the analytical results of human expertise and insights, guiding LLMs to produce outcomes more aligned with human understanding.
\end{enumerate}

\subsection{Intermediate Workspace as Prompt}
\label{sec:prompt}
To steer the LLM summarization, we first extract the representation information from the intermediate workspace, transform it into natural language, and embed it into a prompt.
After reviewing previous visualizations for sensemaking \cite{stasko2007jigsaw, endert2012semantic, davidson2022exploring, keith2023mixed, suh2023sensecape, mahyar2014supporting, andrews2012analyst}, we have identified four typical types of information that can be extracted from the intermediate workspaces for sensemaking tasks, in addition to the standard inputs of documents, background, and expectations provided to LLMs: text-level, insight-level, and structure-level, and connection information. We use JSON format, currently supported by GPT4, to construct prompts that include the four types of information extracted from the intermediate workspace.

\begin{figure}[htb]
  \centering
  \begin{subfigure}[b]{0.25\columnwidth}
  	\centering
  	\includegraphics[width=\textwidth]{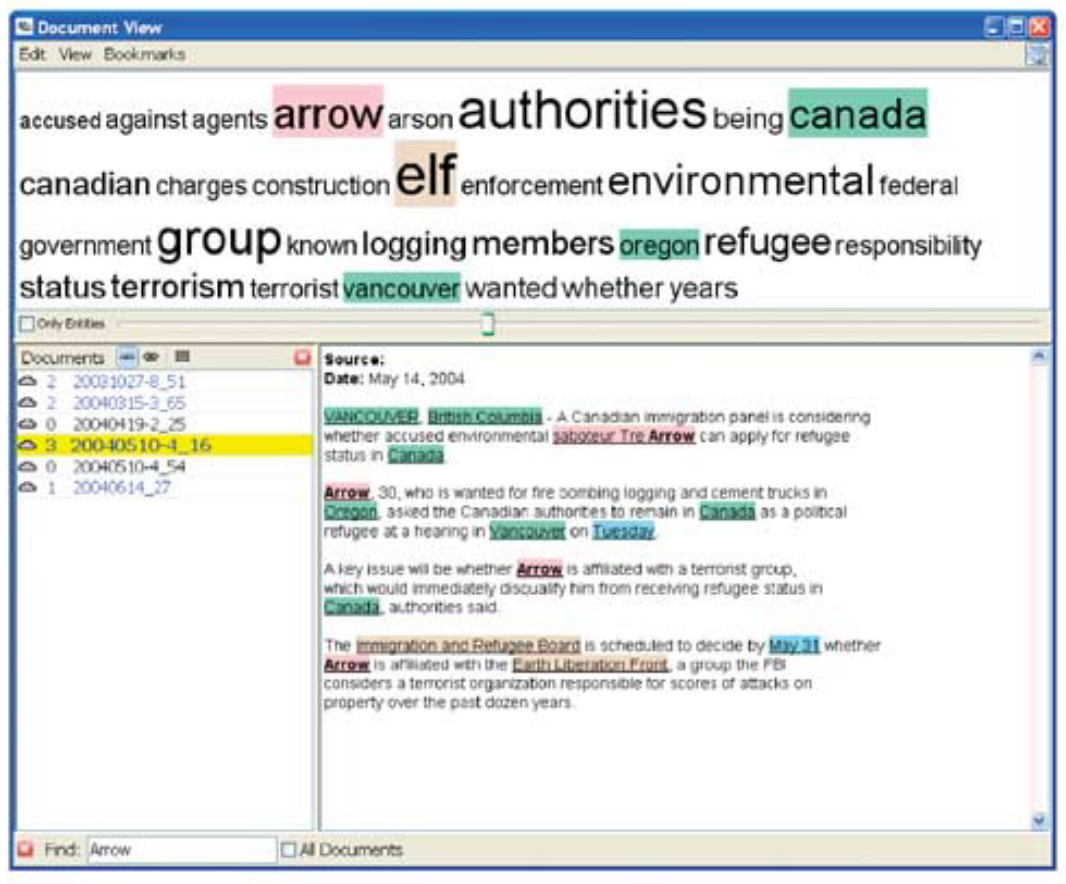}
  	\caption{Highlights}
  \end{subfigure}%
  \hfill%
  \begin{subfigure}[b]{0.2\columnwidth}
  	\centering
  	\includegraphics[width=\textwidth]{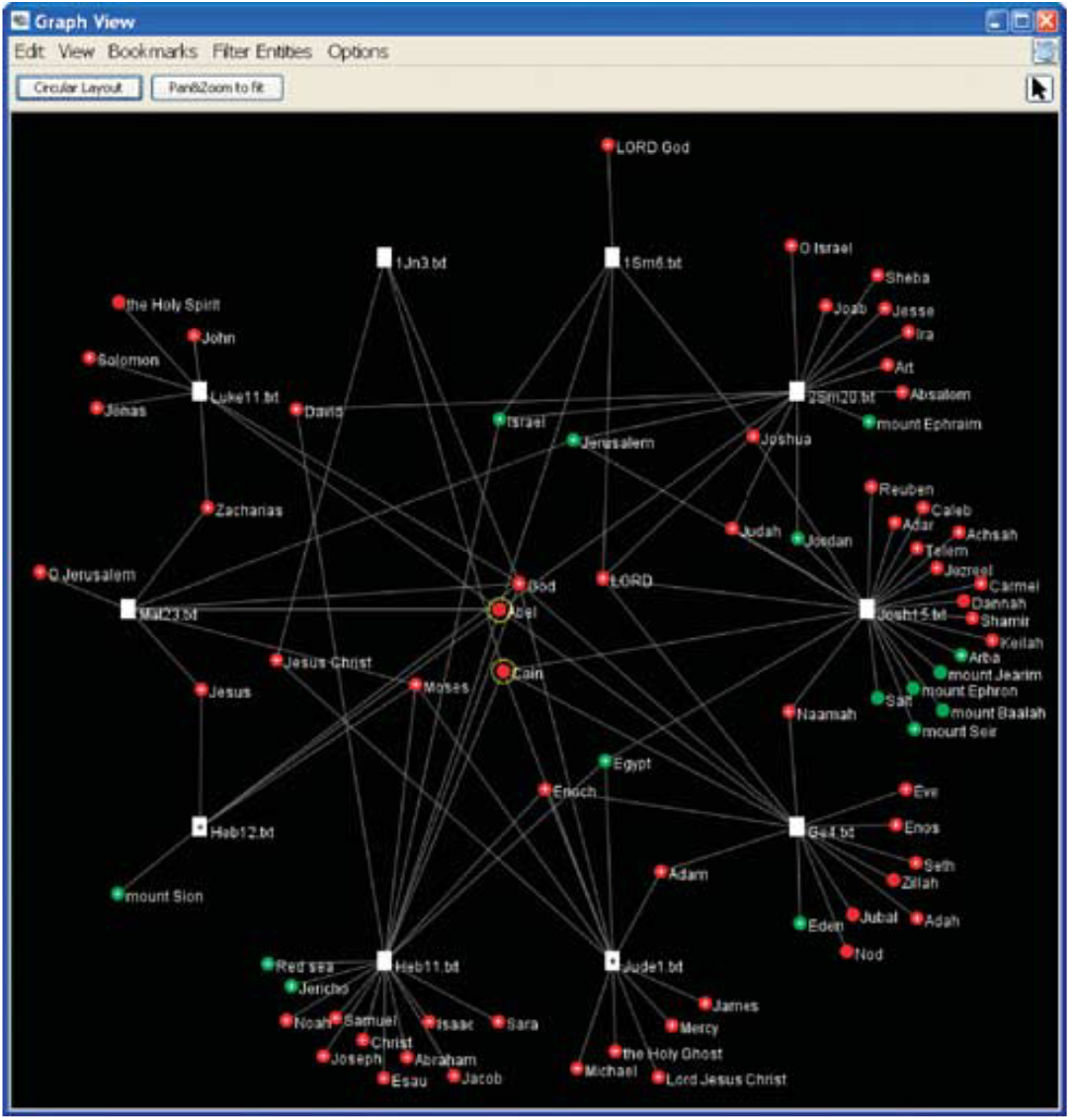}
  	\caption{Text graph}
  \end{subfigure}%
  \hfill
  \begin{subfigure}[b]{0.35\columnwidth}
  	\centering
  	\includegraphics[width=\textwidth]{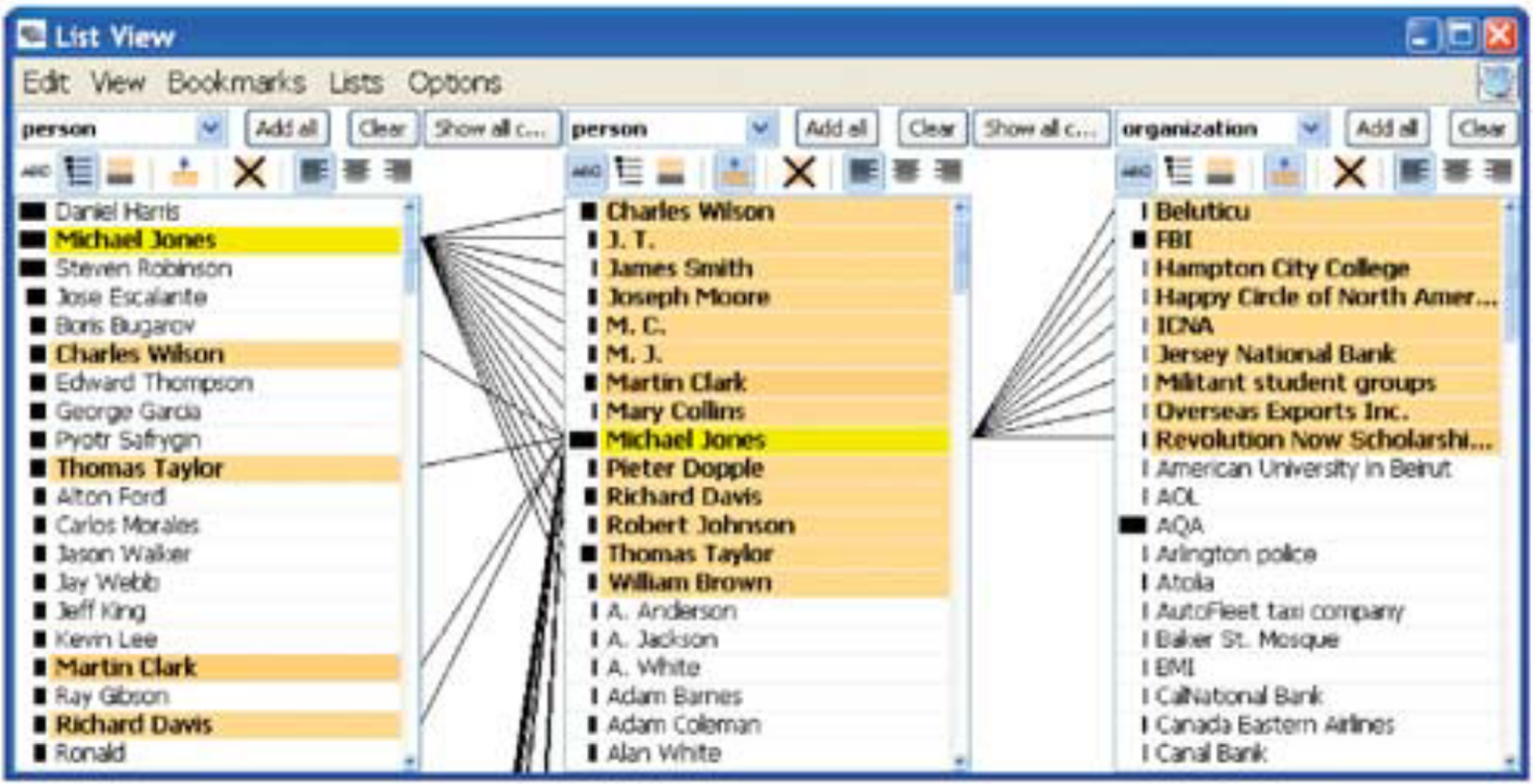}
  	\caption{Text list}
  \end{subfigure}%
  \caption{Visual representations \cite{stasko2007jigsaw} for extracting text-level information.}
  \label{fig:entity}
\end{figure}

\textbf{Text-level information} consists of two parts: the text and the corresponding text weights. Figure \ref{fig:entity} shows various text-level visual representations, such as highlighting, text graphs, and text lists. These visual representations enable analysts to outline essential texts. However, the number of marked texts may vary across different datasets and scenarios. In our prompt engineering, there are two methods. The first method involves directly creating JSON data where the keys are the texts and the values are their calculated importance. The importance weights can be determined based on the frequency of highlighting. For text graphs or text lists, the importance weights can be calculated based on the number of directly connected texts. The second method is suitable for more complex datasets. It involves embedding text-level information within clusters. In this case, the JSON data will be a dictionary where the keys are the cluster names, and the values are the marked texts within those clusters. We use the second method in the experiments.

\textbf{Insight-level information} comprises two parts: the object index and the attached insight.
Humans often record their insights for objects such as documents and clusters in the form of notes and annotations.
In the JSON data for human insights, each key is the object index, while the value is the corresponding insight in natural language. 


\begin{figure}[htb]
  \centering
  \begin{subfigure}[b]{0.5\columnwidth}
  	\centering
  	\includegraphics[width=0.45\textwidth]{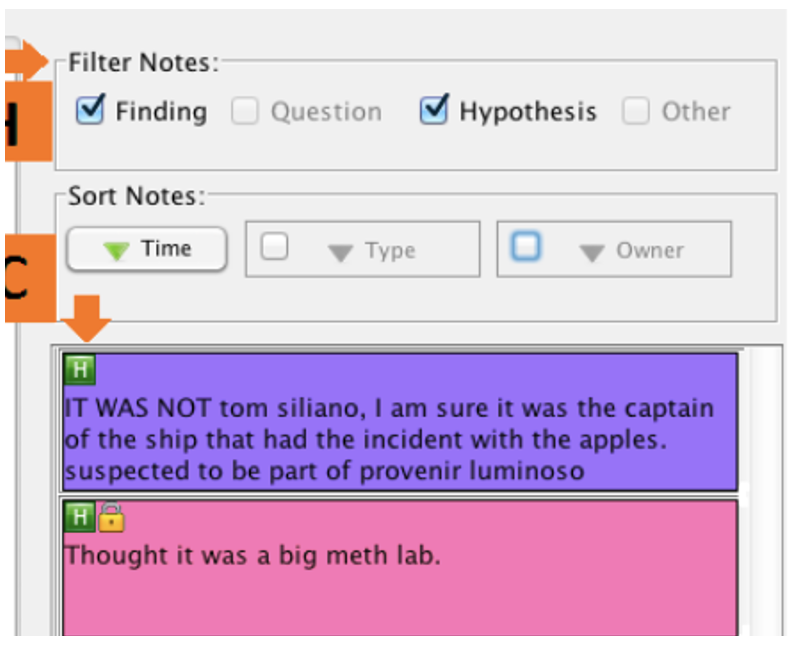}
  	\caption{Notes in list \cite{mahyar2014supporting}.}
  \end{subfigure}%
  \hfill%
  \begin{subfigure}[b]{0.5\columnwidth}
  	\centering
  	\includegraphics[width=0.8\textwidth]{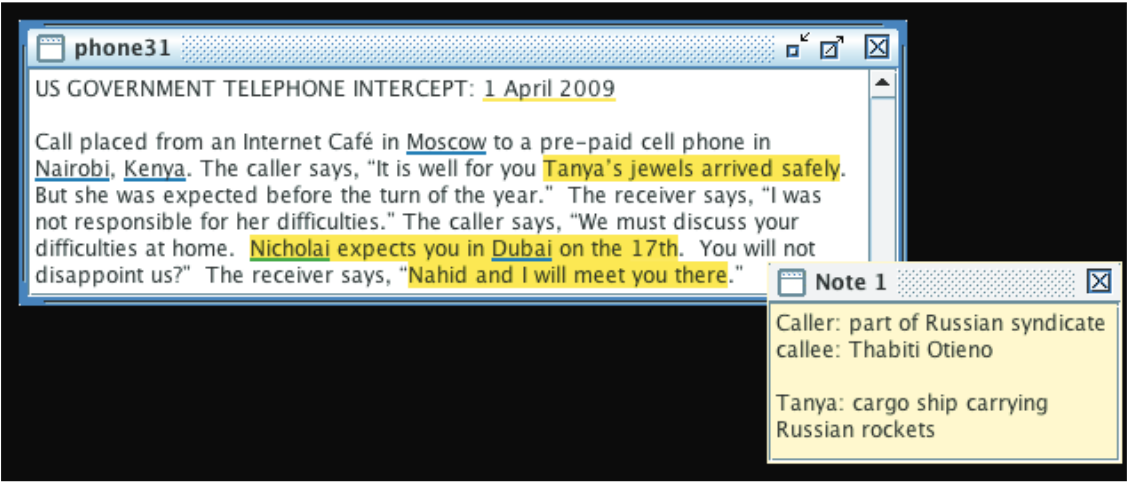}
  	\caption{Notes attached to documents \cite{andrews2012analyst}.}
  \end{subfigure}%
  \caption{Visual representations for extracting insight-level information.}
  \label{fig:document}
\end{figure}

\textbf{Structure-level information} encompasses document clustering information and consists of two parts: cluster name and cluster members (documents). Clusters can be identified implicitly through the proximity of documents or explicitly by grouping or assigning labels. We tried various cluster prompts to prompt clusters and found that incorporating document content with the cluster yields the best results. Thus, in the cluster-level JSON data, each key is the cluster name, and the value is a dictionary containing the included documents' indexes and content. 

\begin{figure}[htb]
  \centering
  \begin{subfigure}[b]{0.3\columnwidth}
  	\centering
  	\includegraphics[width=\textwidth]{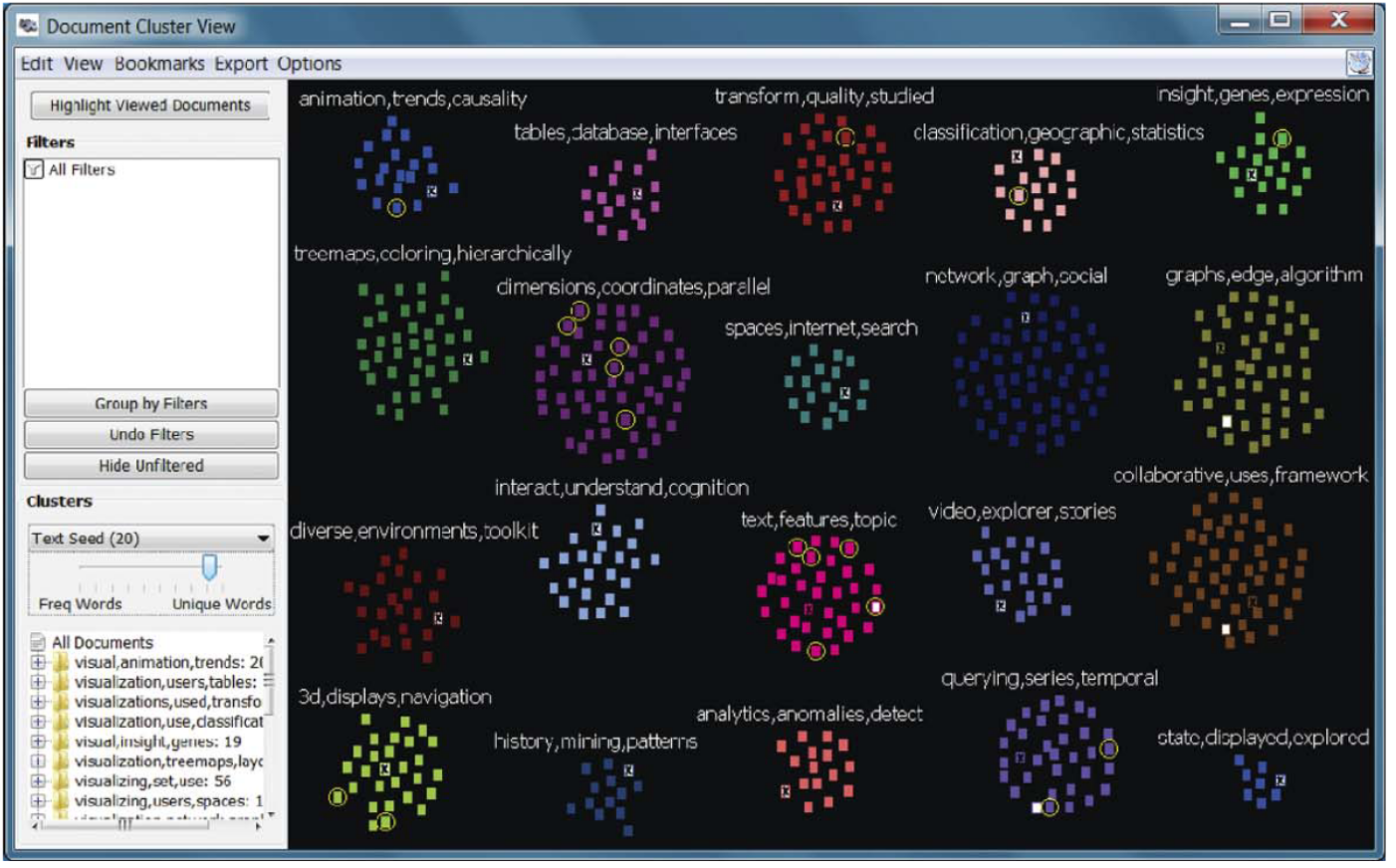}
  	\caption{Clusters are computed and displayed on a new page \cite{gorg2012combining}.}
  \end{subfigure}%
  \hfill%
  \begin{subfigure}[b]{0.3\columnwidth}
  	\centering
  	\includegraphics[width=\textwidth]{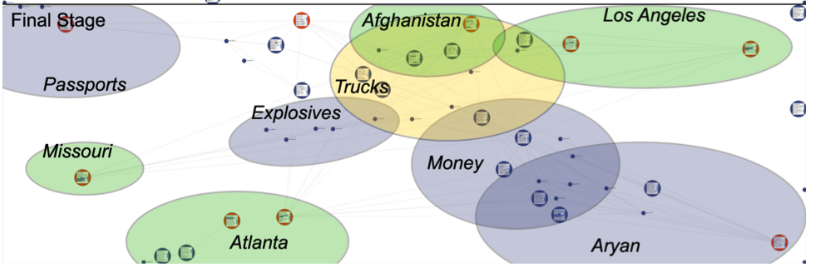}
  	\caption{The documents are clustered by model on the original page \cite{endert2012semantic}.}
  \end{subfigure}%
  \hfill
  \begin{subfigure}[b]{0.3\columnwidth}
  	\centering
  	\includegraphics[width=0.6\textwidth]{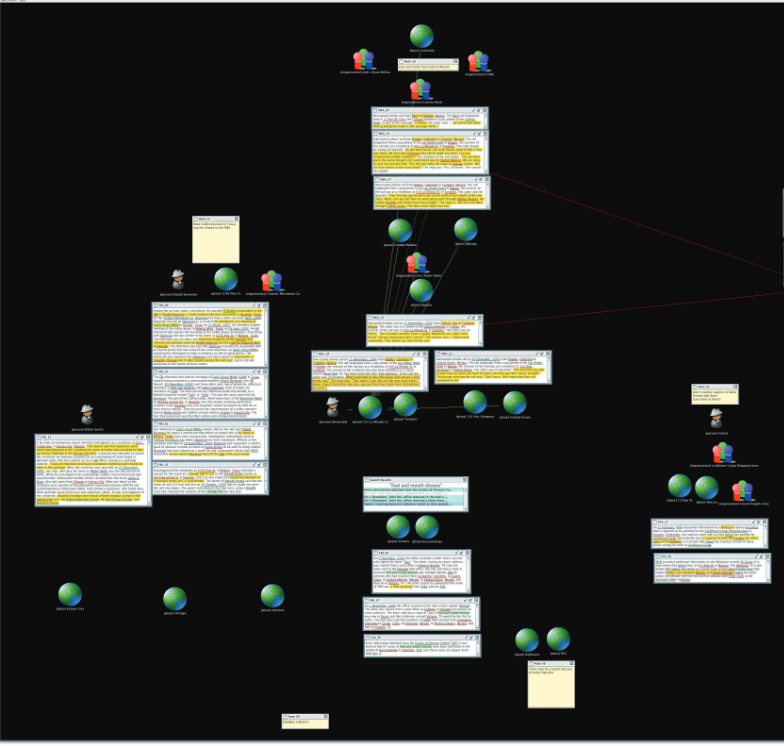}
  	\caption{Documents are manually organized into groups \cite{andrews2012analyst}.}
  \end{subfigure}%
  \caption{Visual representations for extracting structure-level information.}
  \label{fig:structure}
\end{figure}

\textbf{Connection information} represents the relationships among information at the same level, typically depicted as lines or arrows connecting two objects, sometimes with a label. The connected items could be entities, documents, or clusters. We separate connection information because the way we prompt connections differs from the other three levels of information. Connections should include a start object, an end object, and even a label describing the relationship between these two objects.
The connected information JSON data consists of a list of connections. For each connection, the first item represents the start object, the second the end object, and the third the relationship details from marked labels. The item key may correspond to text, document index, or cluster name.

While this list of features is not intended to be exhaustive, we believe it is well-representative enough to enable experimentation as a proof of concept.
An example template prompt can be found in Appendix \ref{apen: prompt1} where we concatenate the different levels of information mentioned above. 

\section{Preliminary Experiment}

We created an intermediate workspace based on ground truth to better understand the enhancements in LLM summarization achieved by integrating it. We then conducted proof-of-concept experiments to assess how the workspace and each type of information impact LLM summarization. We use the GPT-4o model in our experiments.

\subsection{Research Questions}
\begin{itemize}
\item \textbf{RQ1}: Do summarization results generated by the GPT-4o prompted with workspaces align more closely with the ground truth than those generated without them?
\item \textbf{RQ2}: What types of information in the visual workspace help GPT-4o produce summarization that align more closely with the ground truth?
\end{itemize}
\subsection{Methodology}
The dataset we use for this study is a fictional intelligence analysis dataset called \textit{Sign of The Crescent}. It contains 40 documents, of which only 23 documents are relevant. The task of this dataset is to accurately identify the ``\textit{Who}", ``\textit{When}", ``\textit{Where}" accurately, and ``\textit{What}" information for the entire story and all plotlines related to a planned coordinated attack. To test how the workspace influences the LLM summarization results, we created a visual workspace based on the given ground truth of this dataset. 

\subsection{Pipeline}
\label{sm_pipeline}
\begin{figure}[tb]
 \centering 
 \includegraphics[width=\columnwidth]{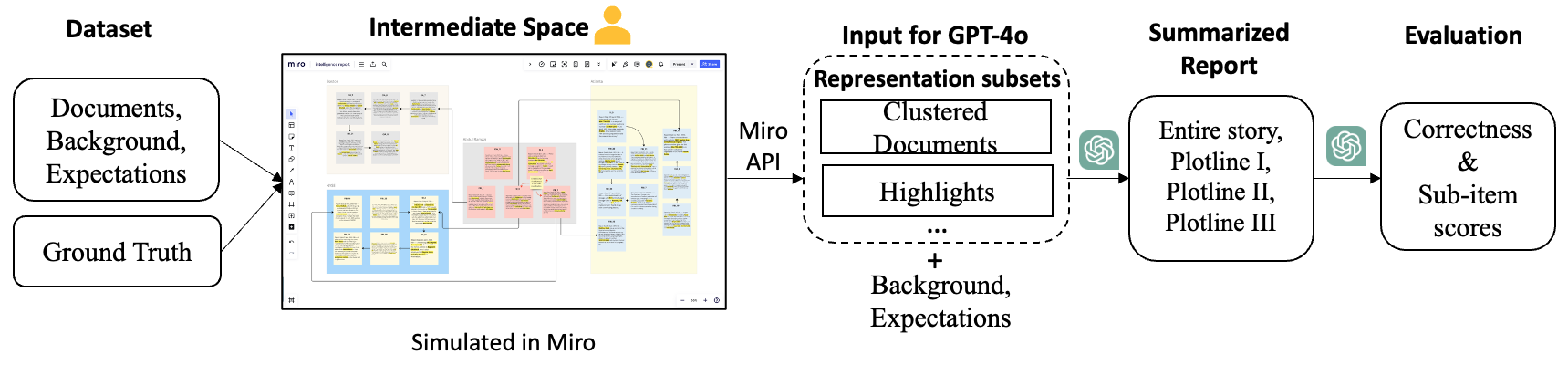}
 \caption{Experiment pipeline}
 \label{fig:pipeline}
\end{figure}
To test how the workspace influences the LLM summarization results, we designed a pipeline (Figure \ref{fig:pipeline}) to evaluate the space-steered sensemaking. 

\begin{figure}[tb]
 \centering
 \includegraphics[width=\columnwidth]{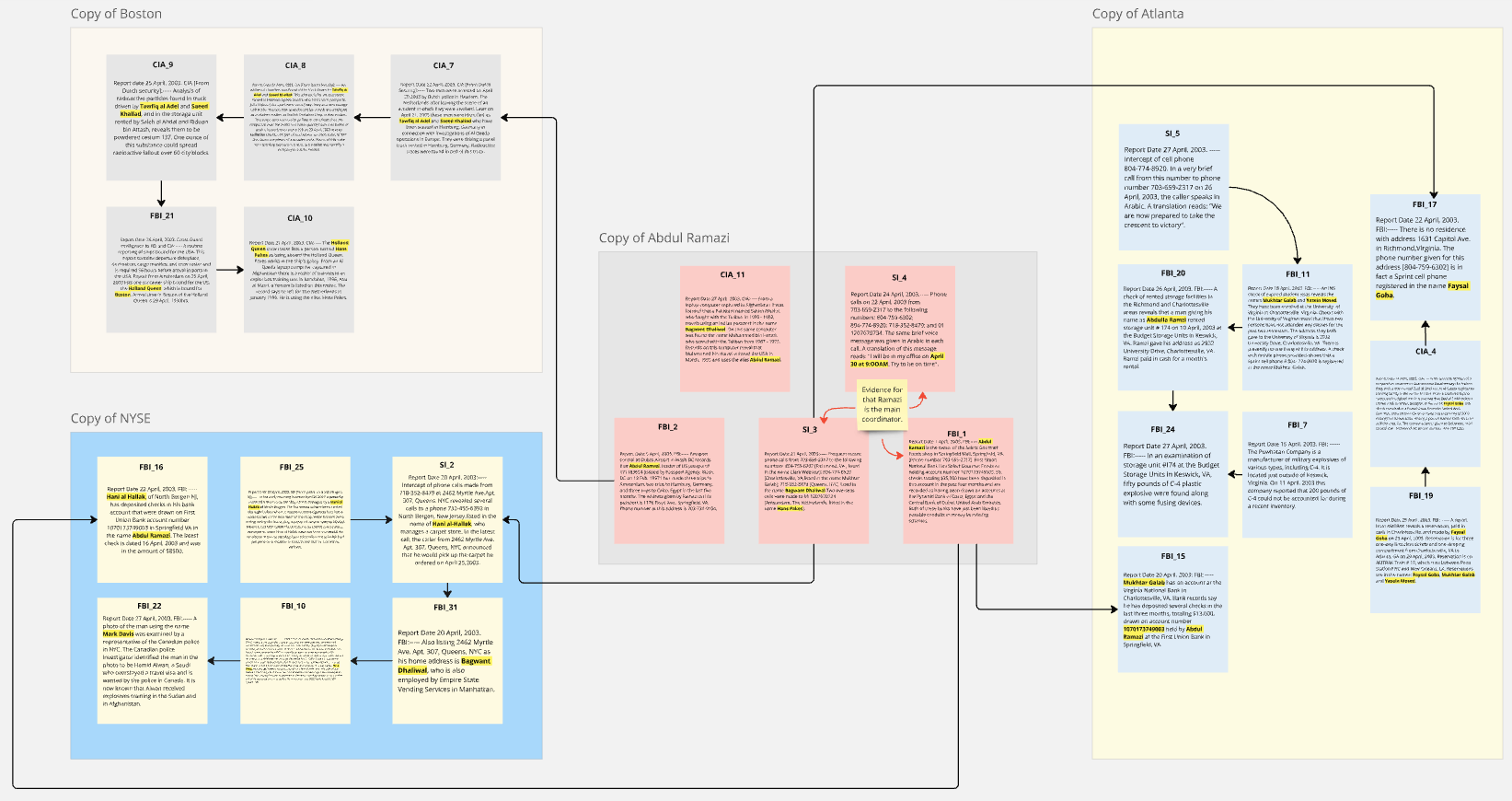}
 \caption{Simulated workspace based on ground truth}
 \label{fig:expspace}
\end{figure}

First, we created a workspace (Figure \ref{fig:expspace}) and the corresponding interactions based on the given ground truth in a document view using Miro\footnote{https://miro.com/}. Miro is an interactive visual workspace that offers essential toolkits and flexible, infinite spaces for human analysis tasks. Based on the four types of information that can be extracted from intermediate workspaces (Section \ref{sec:prompt}), we used corresponding visual representations in the experimented workspace to cover each type. 
To provide text-level information, we highlighted the names of all key participants in the terrorist attack. For insight-level information, we annotated the relevant documents containing details about the central coordinator, directly identifying the central coordinator. We clustered the documents according to the ground truth to provide structure-level information and labeled each cluster topically. For connection information, we connected the relevant documents in a timeline.

We then meticulously collected all visual information in the simulated space using Miro's REST API, ensuring a comprehensive set from which we can form subsets for subsequent experiments.  
We then concatenated a subset (depending on the experimental condition) of the four types of information extracted from the workspace onto the template prompt from Section \ref{sec:prompt}.

Then, we input a prompt (\ref{apen: prompt1}), including the above workspace part, the background information, and expectations into the LLMs and obtained summarized reports. For the experiment, we chose the GPT-4o model for the summarization and the example prompt can be found in Appendix \ref{apen: prompt1}. To ensure the reliability of the results, we replicated the experiment 10 times for each prompt, varying the temperature from 0.1 to 1.0 in increments of 0.1.

Finally, we utilized GPT-4o with the same temperature setting to evaluate the reports consistently. Details of the evaluation are provided in Section \ref{sec:eval}.


\subsection{Evaluation}
\label{sec:eval}
We evaluated the summarized reports using the \textbf{Correctness} rubric from \cite{davidson2022exploring} with a modification: we converted the 5 points of subjective rubrics into fixed objective criteria in Appendix \ref{apen: rubric}. \cite{davidson2022exploring} utilized the same dataset and developed the rubric through a group of intelligence analysis experts. The evaluation of Correctness is based on the ground truth, with a total score of 33 points. The rubrics primarily focus on the plotlines' ``\textit{Who}", ``\textit{When}", ``\textit{Where}", and ``\textit{What}" aspects, evaluating the coverage of these elements in the report. They recruited 8 participants in the user study. The average correctness score of their human summarized reports was 19, with a minimum score of 11 and a maximum score of 29. However, it should be noted that those human participants did not benefit from our ideal simulated human workspace as in our experiments. Yet, they still provided valuable information about human performance on the same task.

To batch assess the correctness of summarized reports, we used the GPT-4o model with a fixed temperature of 0, following the methodology outlined in \cite{galileo2023}: first, we tasked GPT-4o with answering a list of questions about the ``\textit{Who}", ``\textit{When}", ``\textit{Where}", and ``\textit{What}" information in the coordination and other plotlines; then, we had GPT-4o grade its own responses based on the provided rubric. The prompts for questioning and grading can be found in Appendix \ref{apen: prompt2}. For better comparison and understanding, the correctness scores in our results are presented as percentages, calculated by dividing the graded scores by the total points of 33.

\subsection{Results}
\label{sec:results}
\begin{figure*}[tb]
 \centering 
 \includegraphics[width=\linewidth]{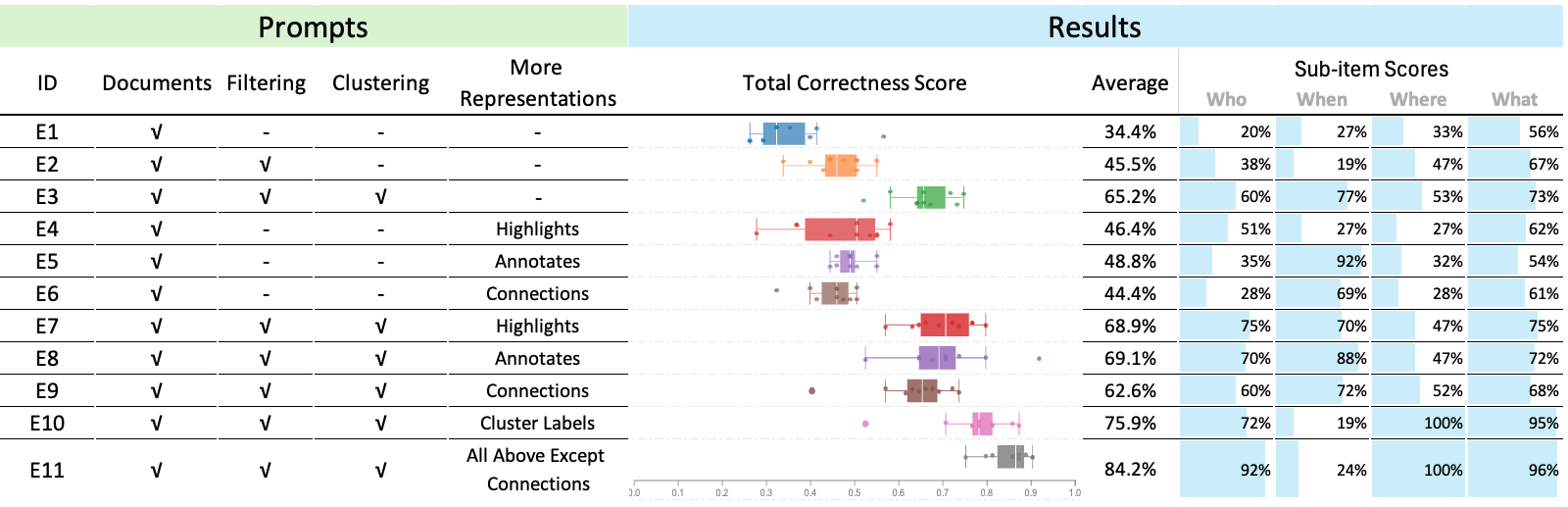}
 \caption{Experiment results. We replicated the experiment for each prompt ten times and visualized the results in a box plot. We also calculated the average correctness percentage rate for each sub-item to gain a deeper understanding of the generated results.}
 \label{fig:result}
\end{figure*}
We conducted experiments on GPT4-generated summarization using various subsets of the visual information from the workspace. The experimental results for both research questions are summarized and presented in Figure \ref{fig:result}.


\subsubsection{RQ1: Without vs. With the intermediate workspace
}
\label{sec:rq1}
We answer RQ1 by comparing the results of \textit{E1, E2, E3, E4, E5}, and \textit{E6}. As discussed in Section \ref{sec:prompt}, we only modified the 
prompted workspace portion in the user part.

In \textit{E1}, where we only included all documents in the prompt, the average correctness for the summarized report is 34.4\%, with most scores falling within the range of (25\%, 42\%). We consider \textit{E1} as the baseline since it does not include any workspace information in the prompt.

Next, we compare the results of \textit{E2} through \textit{E6} with those of \textit{E1} to assess how workspace affects the alignment of LLM-generated summaries with the ground truth.
As the boxplot in Figure \ref{fig:result} shows, filtering the human-selected relevant documents in \textit{E2} improves the average correctness to 45.5\%, with half of the scores falling within (42\%, 52\%). In \textit{E3}, incorporating clustering of the filtered documents, the scores of the generated reports reach an average value of 65.2\%, with half of the scores ranging from 64\% to 73\%. We also added additional representations to the prompts: adding textual highlights (\textit{E4}) yielded scores ranging from 27\% to 58\%; adding annotations (\textit{E5}) concentrated all scores within (42\%, 55\%]; adding connections (\textit{E6}) resulted in most scores falling within (39\%, 52\%).

When input includes the workspace information, GPT-4o tends to generate reports with higher correctness scores than those generated without the workspace. Filtering seems to be a crucial factor, as all experimental prompts--whether through direct filtering (\textit{E2} and \textit{E3}), highlights (\textit{E4}), annotations (\textit{E5}), and connections (\textit{E6})--provide GPT-4o with cues to filter relevant documents.

\subsubsection{RQ2: Types of information in the workspace
}
As shown in Figure \ref{fig:result}, we experimented by prompting GPT-4o with filtering, clustering, highlights, annotations, connections, and cluster names as input.

First, we can observe how a single type of visual representation impacts the generated reports, from \textit{E2} to \textit{E6}. As clustering must be conducted on the filtering result, we incorporate clustering based on filtering. The prompt with the filtering and clustering information yields the best-summarized reports compared to other visual representations. The average scores for filtering, highlights, annotations, and connections are similar, but the score range for highlights spans a 30\% correctness, indicating more significant variability. 

Building on the previous analysis of \textit{E3}, we added more visual representations to filtering and clustering to observe the combined results from \textit{E7} to \textit{E11}. Compared to not adding filtering and clustering information, the correctness scores from  \textit{E7} to \textit{E9} show significant improvement.
Compared to the baseline in \textit{E3}, the average scores for adding highlights (\textit{E7}) and annotations (\textit{E8}) are slightly higher, with a marginal increase in the average sub-item scores for ``\textit{Who}" and ``\textit{What}" information, while connections (\textit{E9}) decrease the scores compared to the baseline. We also attempted to assign names to each cluster (\textit{E10}) based on the ground truth, and the evaluation results indicate that cluster names raised the correctness scores to an average of 75.9\%, with three-quarters of the results scoring higher than 75\%. Based on the above experiments, we combined all extracted visual representations except connections in the new prompt and had GPT-4o summarize reports in \textit{E11}. The average score was 84.2\%, with the highest score reaching 89.4\% (29.5), surpassing the highest score of 29 achieved by a human participant in \cite{davidson2022exploring}. The lowest score in \textit{E11} is 74.2\%, which also exceeds the scores achieved by most human participants in that study.

In summary, providing structural information such as clustering and filtering is most beneficial for LLM summarization; user-specified cluster names help clarify the user's intentions for clustering to the LLMs; highlights and annotations also contribute by providing relevant detailed information in the summarized reports; connections become crucial, especially when filtering and clustering are not employed. The results from \textit{E11} also demonstrate that by effectively manipulating visual representations in the intermediate space, users can guide LLMs to produce aligned summaries. Moreover, with the guidance provided by the intermediate space, LLMs can generate highly accurate summaries comparable to those made by the human participants.


\section{Case Study}
\label{sec:lr}
\begin{figure}[htb]
 \centering 
 \includegraphics[width=0.8\columnwidth]{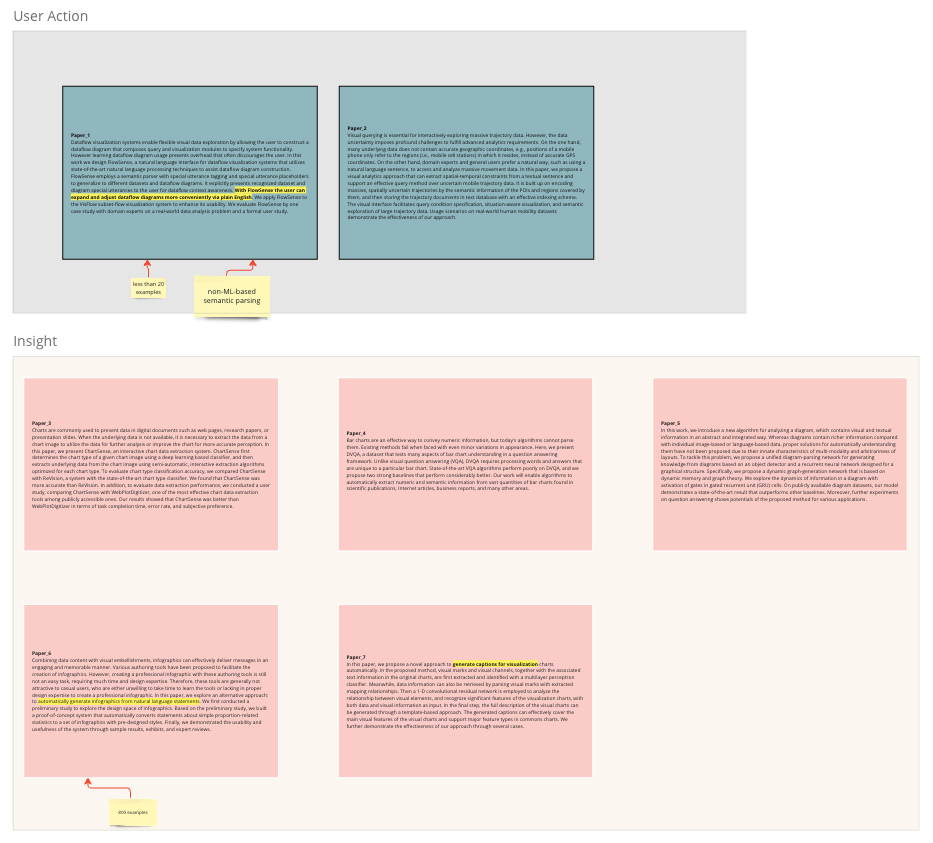}
 \caption{Sample workspace for literature review in Section \ref{sec:lr}.}
 \label{fig:lr}
\end{figure}

We also applied our space-steered method in another area: literature review. For better comparison, we chose a published paper \cite{wang2021survey} to test our method. 
We selected a paragraph from its Section 6.1, which reviews previous methods for training natural language format data for visualization. This paragraph begins with ``\textit{Natural language}" and categorizes the research on training natural language data into two types: ``\textit{user action}" and ``\textit{insight}". For each type, the authors provided example papers. Based on this categorization and the original description, we created a workspace, as shown in Figure \ref{fig:lr}, organizing the example papers into two clusters named after the categories. For each paper in the workspace, we included only its abstract and added an index for it, highlighted the referenced texts from the original description, and annotated the content referred to by the paper but was not present in the abstract. We then generated a prompt from the workspace based on the template prompt from Section \ref{sec:prompt}, as detailed in Appendix \ref{apen: lr_prompt}. Finally, we used GPT-4o to summarize the example papers into a literature review and one generated result is shown in Appendix \ref{apen: lr_result}.

To evaluate whether the workspace enhances the literature review process, we used GPT-4 to summarize the example papers both with and without the workspace, altering only the workspace portion. We observed that the literature review generated with the workspace were significantly better to those without. By incorporating clustering, annotation, and highlighted information into the pre-filtered documents, these visual representations positively influenced the quality of the summarized literature reviews.
\begin{itemize}
\item \textbf{Clustering}. Without providing clustering information, GPT-4o clustered papers differently each time, despite running the same prompt three times with the same temperature setting. For instance, the paper \cite{yu2019flowsense}, which was originally categorized under ``\textit{user actions}", GPT-4o variably classified it as ``\textit{automatic Content generation}," ``\textit{visual querying and semantic exploration}," and ``\textit{natural language interfaces for visualization systems}". With clustering information, duplicated tests showed that GPT-4o consistantly categorized the example papers into the two given types and summarized each type accordingly. Additionally, GPT-4o was able to interpret the meaning of the cluster names such as ``\textit{user actions}" and summarize the clustered papers from the perspective of these cluster names. 
Thus, incorporating clustering information in the workspace allows researchers to steer the LLM summarization for literature reviews.
\item \textbf{Annotations}. Annotations offer additional information beyond the provided texts.
We included only abstracts in the prompt, as they are highly condensed introductions to the work. However, researchers often want to discuss additional details. They can annotate this information in the workspace, and it will be summarized into the literature review. 
Interestingly, GPT-4o selectively adopts annotations. For instance, one example paper labeled less than 20 examples, and despite multiple attempts, GPT-4o consistently refused to include it. Researchers may need to expand on their annotations to clarify why certain annotations are important.
\item \textbf{Highlights}. With highlights, GPT-4o includes the highlighted information prominently in its summarization. For example, in the case of ``\textit{AutoCaption}" \cite{liu2020autocaption}, we annotated it as ``\textit{generate captions for visualization}". When summarizing without the workspace, GPT-4o sometimes used alternative phrases like ``\textit{descriptive text}," whereas in the space-steered summarization, it consistently used the original term ``\textit{captions}". Highlights are particularly useful when researchers aim to retain specific expressions from cited papers.
\end{itemize}

We also compared the space-steered summarized literature review with the original description. Even though output requirements constrain LLM-generated literature reviews and are not as flexible as human-written ones, they closely align with the original writing. The GPT-4o model also divides the papers into the same two categories and the interpretations of ``\textit{user actions}" and ``\textit{insight}" match the texts in the original paper.
 Additionally, LLM-generated paper summarizations fit well with human-written summaries. For example, for \cite{liu2020autocaption}, the original writing is ``\textit{AutoCaption generates a text caption to describe the insights of a visualization based on four types of chart features detected.}" At the same time, GPT-4 writes, ``(\textit{AutoCaption) focuses on generating captions for visualization charts, combining data extraction with visual feature recognition to produce comprehensive summaries of visual data.}"


\section{Discussion}

\subsection{Four types of Information}
We utilized GPT-4 to summarize the sensemaking tasks by prompting four types of information in Section \ref{sec:prompt}. The four identified types of information in the intermediate workspace can support the sensemaking process described by Pirolli and Card \cite{pirolli2005sensemaking}, which consists of two primary cycles: the foraging loop involves information seeking, searching and filtering, and the sensemaking loop involves iterative development of schemas that best fit the evidence. Future work can expand these methods to transform various types of intermediate workspaces into prompts.

Human analysts externalize their cognitive process in the intermediate space by creating the four types of information. They use text-level interactions like highlights to mark and extract detailed information, annotations to add insights and make notes, logically connect relevant documents with lines or arrows, and structure information to externalize their mental models. The intermediate space enables human analysts to offload their cognitive processes and structure the relevant information for complex sensemaking tasks.

Interestingly, our preliminary experiment results show that the four types of information are essential for both humans and GPT-4. Text-level information helps LLMs decide which details to focus on, insight-level information guides LLMs to find evidence supporting human insights, structure-level information eases the LLM analysis process by pre-clustering relevant documents, cluster names indicate how humans have clustered those documents, and connected information links the documents that humans consider relevant. The effective combination of those types of information can guide LLMs to generate summaries that closely aligned with human intent. It can save analysts much time and effort compared to manually writing the summaries.


\subsection{Using LLMs for Summarization}
In our experiments, directly inputting all documents resulted in an average correctness score of only 34.4\%. Instead of finding interrelated documents and connecting all the facts into a complete story, GPT-4 tends to identify different patterns of events without proper filtering and clustering. However, when information extracted from the intermediate workspace--especially filtering and clustering information--is included in the prompts, the correctness scores of LLM-generated summaries significantly increase.
This suggests LLM summarization in cognitively intensive sensemaking tasks can significantly benefit from human visual workspaces. Moreover, current LLM chat interfaces pose challenges for the iterative interactions needed for sensemaking tasks. Users often need to scroll up and down to read documents, recheck specific information from previous LLM responses, and find additional space, such as blank paper, to offload their mental models. Such interactions can be significantly streamlined in the intermediate workspace to be more aligned with cognitive models.

Although workspaces provide visual representations to facilitate human sensemaking, there are still some challenges in using LLMs for summarization. First, how can we enable users to manipulate workspaces to steer LLM summarization effectively? What is the learning curve? Second, it is laborious for humans to compare the workspace with the summarization after each generation. Users must check the summarized reports line by line to ensure they include all critical content and insights and do not contain irrelevant information or incorrect speculations. Our ongoing project aims to implement a system and conduct a user study to address these questions and resolve the challenges.






\subsection{Intermediate Workspace as a Medium for Human-AI Collaboration}
As Section \ref{sec:rq1} discussed, the simulated intermediate workspace can steer LLM summarization results more aligned with the ground truth. The visual representations in the intermediate space provide the text, insight, connection, and structure levels of information to guide LLM summarization. The intermediate workspace also makes communications between humans and LLMs more efficient. The two primary functions of the intermediate workspace, external memory, and semantic layer, can provide significant context for the LLM summarization.

In our experiments, one round of LLM summarization with the intermediate space takes around 10 seconds. Still, the same task for human summarization in \cite{davidson2022exploring} took around one hour, while the best reports generated by both GPT-4 and humans achieved similar correctness scores. Therefore, the intermediate workspace can serve as the medium for human-AI collaboration: it significantly accelerates the summarization process for humans while guiding LLMs to align their summarization more closely with human intent.

\begin{figure}[htb]
 \centering 
 \includegraphics[width=0.8\columnwidth]{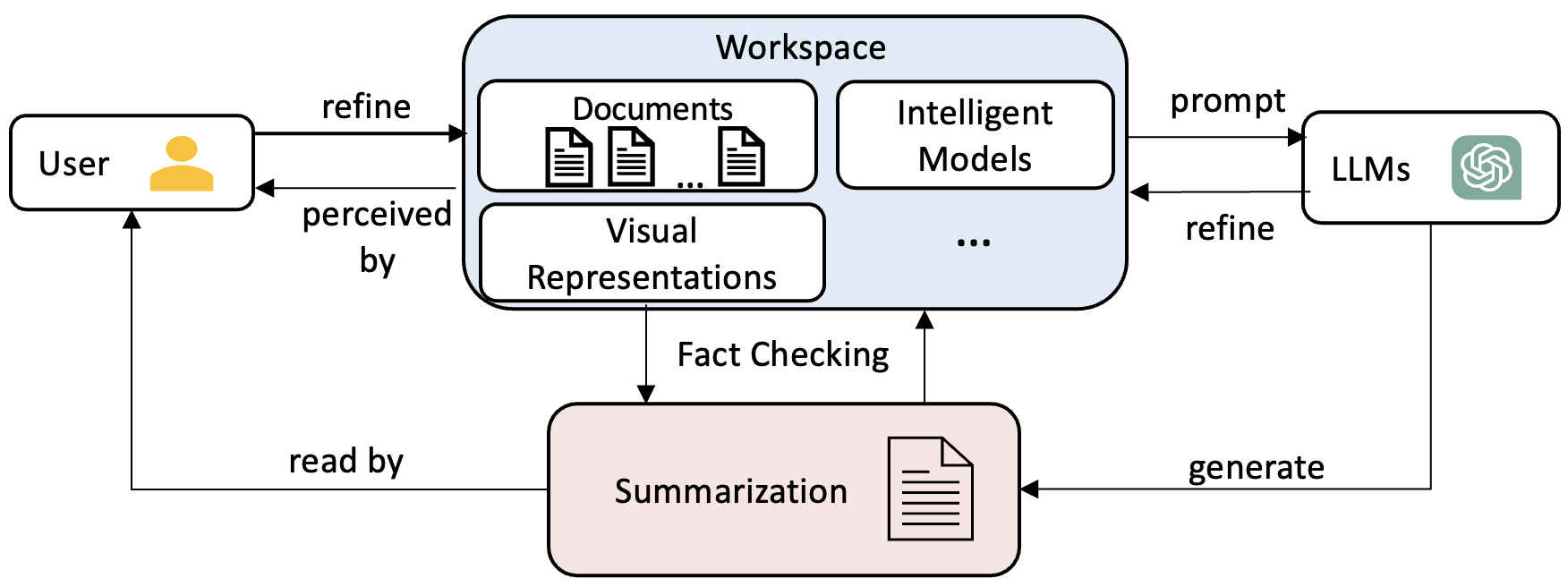}
 \caption{Workspace as a medium for human-AI collaboration for summarization.}
 \label{fig:future}
\end{figure}

This medium for human-AI collaboration presents attractive future research opportunities (Figure \ref{fig:future}). In addition to steering, the intermediate workspace can be a mixed-initiative space by incorporating intelligent models to assist in constructing the workspace and aid the human sensemaking process. It could lead to further automation and steering opportunities in the summarization process. It could also serve as a medium to visualize connections from source material to the summary, thus enabling verification of LLM summarization. Also, it could serve as an interactive method for the iterative refinement of summaries.

Furthermore, our results indicate a potential congruence between cognitive and AI sensemaking processes. Cognitive Space to Think and LLM context spaces serve similar purposes in their respective agents. This congruence could be exploited to improve human-AI collaborative sensemaking in the future.



\section{Conclusion and Future work}
We propose space-steered summarization, where newly-introduced human-created 
workspaces guide LLM summarization in sensemaking tasks. This workspace serves as an intermediate step for LLM summarization. 
As an initial attempt to integrate visual workspaces for steering LLM summarization, we conduct a series of experiments to evaluate the impact of the visual workspace on summarization and assess the visually-steered method's effectiveness. The results show that the workspace can readily steer an LLM to be aligned with the desired outcome, indicating that a well-constructed workspace can enable LLMs to generate accurate summaries for user sensemaking tasks.  
In terms of sensemaking applications, this allows analysts to steer the LLMs toward useful results. Of course, this method depends on the LLMs being given good-quality workspaces generated by the human analysts.


The effectiveness of the intermediate workspace in enhancing summarization alignment with human intent through visual workspace is tested through experiments. The results confirm that the intermediate workspace significantly improves LLM summarization by providing additional structural context and augmenting their external memory. It sets the foundation for future work on building interactive systems by applying the lessons learned in this paper. We plan to conduct user studies to understand how such interactive systems can help analysts and AI collaborate to accomplish complex summarization tasks efficiently.

\bibliographystyle{abbrv-doi}

\bibliography{template}

\newpage
\appendix
\section{Appendix}
\subsection{Prompt for Summarization}
\label{apen: prompt1}
We present an example of a prompt we used for GPT-4o summarization. 

\textbf{System:}

Imagine you are a FBI agent who is analyzing the related events and your basic task is to predict the nature of the terrorists' threat including when and where this threat will be carried out.

\textbf{Assistant:}

First paragraph: summarizing your findings from those documents as the first paragraph of the Bottom Line Up Front manner. It should include the necessary information of Who, When, Where, and What.

Middle paragraphs(must include every sub point):

Outline Point 1 (SUB\_POINT\_NAME): [SUMMARIZATION first in a sentence] [ANALYSIS CONTENT must include specific information of Who, When, Where and What. Please write a report rather than list all the points.]

Outline Point 2 (SUB\_POINT\_NAME): [SUMMARIZATION first in a sentence]...

...

Outline Point N (SUB\_POINT\_NAME): [SUMMARIZATION first in a sentence]...

Last paragraph: draw conclusions of how those sub points connect to each other, inferences, and speculate on causes and effects. Do not re-list the events mentioned here.

\textbf{User message part I: clustered filtered documents}

\{

  'NYSE':\{
  
    \hspace*{2cm}’SI\_2’: 'Report Date 20 April. 2003:----- lntercept of phone calls made from 718-352-8479 at 2462 Myrtle Ave.Apt. 307, Queens. NYC revealed several calls to a phone 732-455-6392 in North Bergen, New Jersey.listed in the name of Hani al\-Hallak, who manages a carpet store. In the latest call, the caller from 2462 Myrtle Ave. Apt. 307, Queens, NYC announced that he would pick up the carpet he ordered on April 25,2003.’,
    
    \hspace*{2cm}... (other documents)
    
    \hspace*{1cm}\},
    
    ... (other clusters)
    
\}

\textbf{User message part II: visual representations}

I have attached annotations to the nodes. You should generate the report based on the annotation information I provided.

Annotation information:

[

	\hspace*{1cm} \{
 
		\hspace*{2cm} node: "FBI\_1",
  
		\hspace*{2cm} annotation: "Evidence for that Ramazi is the main coordinator"
  
	\hspace*{1cm} \},
 
    \hspace*{1cm}... (annotations in other clusters)
 
]

I have some word weights of important words. Please add the details of those words in the report.

Words and weights:

\{

  \hspace*{1cm}"cluster1":\{
  
  
    \hspace*{2cm}'Hani al-Hallak': 2,
    
    \hspace*{2cm}'Mark Davis,': 1,
    
    \hspace*{2cm}'Bagwant Dhaliwal': 1,
    
  \hspace*{1cm}\},

  \hspace*{1cm} ...(Highlights in other clusters)
  
\}

\subsection{An Example of Space-steered Summarization Results}
\textbf{Bottom Line Up Front: }
An analysis of intercepted communications, financial transactions, and intelligence reports indicates a coordinated terrorist threat involving multiple individuals, set to be carried out on April 29, 2003. Key suspects include Hamid Alwan (alias Mark Davis), Bagwant Dhaliwal (alias Sahim Albakri), Hani al-Hallak, Abdul Ramazi (alias Muhammed bin Harazi), Steven Clark, Shiela Watson, Mukhtar Galab, Yasein Mosed, Faysal Goba, Tawfiq al Adel, Saeed Khallad, and Hans Pakes. Potential targets are the New York Stock Exchange, Boston Harbor, and possibly an Amtrak train traveling from Charlottesville, VA, to Atlanta, GA. The nature of the threats suggests the use of explosives and possibly radioactive material.

\textbf{Details and Analysis:}

\textbf{New York Stock Exchange:}
Intercepted communications and background checks have highlighted several anomalies among individuals with access to the NYSE. Steven Clark, Shiela Watson, and Mark Davis have questionable backgrounds and, specifically, Mark Davis, identified as Hamid Alwan, is trained in explosives. Alwan is connected to Bagwant Dhaliwal, who also resides at 2462 Myrtle Ave. Apt. 307, Queens, NYC—a location central to multiple suspicious activities. Given the presence of these individuals within the NYSE and their potential for carrying out coordinated actions, there is a high risk of an attack targeting this financial institution.

\textbf{Boston Harbor: }
The container ship Holland Queen, departing from Amsterdam and bound for Boston Harbor on April 29, carries significant risk. Among the crew is Hans Pakes, identified as Abu al Masri, trained in explosives. Recent Dutch intelligence reports arrest of two Al Qaeda operatives in Haarlem with radioactive traces and TNT. This suggests that the Holland Queen could be used to transport a radiological dispersal device (RDD or "dirty bomb") to the harbor. The arrival and explosive capabilities indicate a potential attack aimed at causing extensive damage and mass panic.

\textbf{Atlanta Amtrak: }
The intercepted cell phone calls, expired student visas, and financial transactions between Charlottesville and Springfield linked to Abdul Ramazi’s account suggest a planned attack involving explosives. Faysal Goba, Mukhtar Galab, and Yasein Mosed have reservations on an Amtrak train from Charlottesville to Atlanta on April 29, 2003. Given Goba's known explosives training and the recovered C-4, it is likely that this group plans to detonate explosives on the train, potentially causing significant casualties and disruption.

\textbf{Connections, Inferences, and Speculation:}
This analysis uncovers a web of interlinking individuals and activities pointing towards a multi-faceted terrorist plot orchestrated by operatives trained in explosives. Financial ties, shared addresses, and overlapping networks suggest a coordinated effort. The timing—April 29—and geographical spread—New York, Boston, Atlanta—indicate synchronized attacks aimed at high-profile targets, leading to widespread chaos. The involvement of radiological materials heightens the potential impact, especially at Boston Harbor. This coordinated effort likely leverages the individuals' diverse access points to inflict maximum damage simultaneously, disrupting financial, transportation, and public safety systems.

In conclusion, the threats are imminent and need immediate counter-terrorism measures to prevent potentially devastating attacks on U.S. soil. The FBI, Homeland Security, and local authorities must collaborate closely to secure the identified targets and neutralize the operatives involved.

\subsection{Prompts for Evaluation}
\label{apen: prompt2}

\subsubsection{First Prompt}

\hspace*{0.35cm}\textbf{System:}

Please read the report first, then answer the following questions.

\textbf{Assistant:}

Here are the question I want you to answer:

Q1. Does it identify coordinated terrorist action is planned to take place?

Q2. Does it identify this terrorist action is planned to take place?

Q3. Does it identify the coordinator? Who is the coordinator?

... (other questions)

\subsubsection{Second Prompt}

\textbf{System:}

Please grade based on the given answers according to the following rubrics and return in JSON:

\textbf{Assistant:}

Please grade the report according to the following rubrics:

Q1. identify coordinated terrorist action is planned to take place.

* Failure to identify a coordinated action - 0

* Somewhat identifies a coordinated action - 1.5

* Correctly identifies a coordinated action - 2.5

... (Other questions and rubrics)

\subsection{Rubrics}
\label{apen: rubric}

We have included the same rubrics in \cite{davidson2022exploring} here, with the modification of 5), as the previous subjective rubrics were not suitable for GPT-4o to use for grading.

1) who:

 \hspace*{0.5cm} a) Plotline 1 - 3 Actors (1 point each)

 \hspace*{0.5cm} b) Plotline 2 - 3 Actors (1 point each)

  \hspace*{0.5cm} c) Plotline 3 - 3 Actors (1 point each)

   \hspace*{0.5cm} d) Coordinator Actor (3 points)

2) What:

    \hspace*{0.5cm}a) Coordination Action (5 points)

3) When:

    \hspace*{0.5cm}a) Date (3 points)

    \hspace*{0.5cm}b) Time (2 points)

4) Where:

    \hspace*{0.5cm}a) Plotline 1 Location (2 points)
    
    \hspace*{0.5cm}b) Plotline 2 Location (2 points)
    
    \hspace*{0.5cm}c) Plotline 3 Location (2 points)

5) Other

    \hspace*{0.5cm}The previous rubrics were abstract, requiring the grader to assign points based on the accuracy of information and claims. We have modified this section to be more objective with fixed rubrics:
    
    \hspace*{0.5cm}a) Correct explosive items are mentioned in Plotline 1 (1 points)

    \hspace*{0.5cm}b) Correct explosive items are mentioned in Plotline 2 (1 points)

    \hspace*{0.5cm}c) Correct explosive items are mentioned in Plotline 3 (1 points)

    \hspace*{0.5cm}d) Not include the locations except the ground truth (2 points)

\subsection{Prompt for Literature Review}
\label{apen: lr_prompt}
\textbf{System:}

I am writing a literature review on natural language processing used to solve visualization problems. I have clustered the papers for you in JSON form. Please generate a literature review based on the provided clusters and interactions. Only generate one paragraph for one cluster, which means the analysis for one paper can only take at most two sentences. You must refrain from simply summarizing the papers.

\textbf{Assistant:}

You should write the literature review based on this template:

Summary: summarize your findings as the first part of the literature review. If you refer to one document, please cite it by [paper id].

Cluster summarization: [Summarize the cluster in one sentence first.][Please elaborate your analysis in detail. Remember to cite the document you refer to using [document id]. Do not generate the summarization for each paper and only generate one paragraph for one cluster.][Conclude by comparing the methods of the papers in the technique and share your insights.]

Conclusion: conclude your findings as the final part of the literature review.

\textbf{User message part I: clustered filtered papers}

\{

  \hspace*{1cm}"User Action": \{
  
  \hspace*{2cm} "paper\_1": "Dataflow visualization systems enable flexible visual data exploration by allowing the user to construct a ... (Omit other texts in the abstract here.)",
  
  \hspace*{2cm} ... (other papers)
  
  \hspace*{1cm}\},
  
  ... (other clusters)
  
\}

    



    
    
    
    

\textbf{User message part II: visual representations}

I highlighted the following words followed by the highlighted frequency, please add them to the literature review.

Highlight information:

\{

\hspace*{1cm} “With FlowSense the user can expand and adjust dataflow diagrams more conveniently via plain English”: 1,

\hspace*{1cm} “automatically generate infographics from natural language statements”: 1,

\hspace*{1cm} “generate captions for visualization”: 1

\}

\vspace{10pt}
I have annotations tied with papers. You should generate the literature review according to the annotation information I provided.

Annotation information:

[

 \hspace*{1cm} \{
 
 \hspace*{2cm} node: “paper\_1”,
 
  \hspace*{2cm} annotation: “20 examples“,
  
 \hspace*{1cm} \},

 \hspace*{1cm} \{
 
 \hspace*{2cm} node: “paper\_1”,
 
  \hspace*{2cm} annotation: “non-ML-based semantic parsing“,
  
 \hspace*{1cm} \}, 
 
 \hspace*{1cm} \{
 
  \hspace*{2cm} node: “paper\_6",
  
  \hspace*{2cm} annotation: "800 examples"
  
 \hspace*{1cm} \}
 
]

\subsection{Literature Review Result}
\label{apen: lr_result}

\textbf{Summary:} The application of natural language processing (NLP) in addressing visualization problems emphasizes the enhanced usability of data interaction systems by leveraging plain English commands [paper\_1], automated creation of infographics [paper\_6], and generating captions for visualization [paper\_7]. These efforts seek to bridge the gap between raw data, visual representation, and user interaction, making complex data manipulation more accessible to non-experts.

\textbf{User Action Cluster:} This cluster focuses on the integration of NLP to reduce the learning curve and enhance user interactions with data visualization systems. "FlowSense: A Natural Language Interface for Visual Data Exploration within a Dataflow System" [paper\_1] addresses the complexity of dataflow visualization systems by enabling users to manipulate dataflow diagrams using natural language commands, thereby facilitating interaction without a steep learning curve. The system's non-ML-based semantic parsing with 20 examples demonstrates robustness in generalizing to various datasets, ensuring ease of use. Similarly, "A natural-language-based visual query approach of uncertain human trajectories" extends this concept to querying spatially uncertain trajectory data through a visual interface, highlighting the challenges of handling imperfect data while still enabling effective interaction using natural language. 

\textbf{Insight Cluster:} This cluster underscores methodologies for automatic data extraction and comprehension from various types of visual representations using NLP. "Chartsense: Interactive data extraction from chart images" [paper\_3] presents a system for extracting data from chart images, leveraging deep learning for chart type classification and improving task efficiency compared to existing tools. Next, "DVQA: Understanding data visualizations via question answering" [paper\_4] introduces a dataset designed to test bar chart comprehension in a question-answering framework, pushing the boundaries of semantic and numeric information extraction from visual data. Furthermore, "Dynamic graph generation network: Generating relational knowledge from diagrams" [paper\_5] proposes an advanced network to parse diagrams into knowledge structures, employing graph theory and recurrent networks to interpret multi-modal, graphical data. This approach reveals the intricate dynamics of diagram information processing and positions it as a state-of-the-art method. Another significant development comes from "Text-to-Viz: Automatic generation of infographics from proportion-related natural language statements" [paper\_6], which employs NLP to automatically generate infographics, demonstrating usability with 800 examples. Finally, "AutoCaption: An approach to generate natural language description from visualization automatically" [paper\_7] focuses on generating captions for visualization charts, combining data extraction with visual feature recognition to produce comprehensive summaries of visual data. This method ensures that key visual elements are effectively communicated to the user.

\textbf{Conclusion:} Utilizing NLP for solving visualization problems enhances user engagement and simplifies complex data interaction. These approaches range from enabling plaintext dataflow adjustments with FlowSense [paper\_1] to creating infographics automatically from natural language statements [paper\_6], and generating captions for visualizations [paper\_7]. Each contribution underscores the potential of NLP to democratize data visualization, making sophisticated analytical tools accessible to users with varying levels of expertise. The innovative methodologies across these clusters suggest that as NLP technology continues to evolve, its integration with data visualization will further break down barriers to data comprehension and analysis.

\clearpage
\end{document}